# Modeling water radiolysis with Geant4-DNA: Impact of the temporal structure of the irradiation pulse under oxygen conditions

Tuan Anh Le[1], Hoang Ngoc Tran[2*], Serena Fattori[3], Viet Cuong Phan[4], Sebastien Incerti[2]

[1] Institute for Nuclear Science and Technology, VINATOM, 179 Hoang Quoc Viet, Hanoi, Vietnam

[2] Université de Bordeaux, CNRS, LP2I, UMR 5797, F-33170 Gradignan, France

[3] Istituto Nazionale di Fisica Nucleare (INFN), Laboratori Nazionali del Sud (LNS), Catania, Italy

[4] Hanoi Irradiation Center, VINATOM, 2PXV+473, QL32, Minh Khai, Hanoi, Vietnam

*tran@lp2ib.in2p3.fr

**Abstract**

The differences in $H_2O_2$ production between conventional (CONV) and ultra-high dose rate (UHDR) irradiations in water radiolysis are still not fully understood. The lower levels of this radiolytic species, as a critical end product of water radiolysis, are particularly relevant for investigating the connection between the high-density energy deposition during short-duration physical events (ionizations or excitations) and biological responses of the FLASH effect. In this study, we developed a new Geant4-DNA chemistry model to simulate radiolysis considering the time structure of the irradiation pulse at different absorbed doses to liquid water of 0.01, 0.1, 1, and 2 Gy under 1 MeV electron irradiation. The model allows the description of the beam's temporal structure, including the pulse duration, the pulse repetition frequency, and the pulse amplitude for the different beam irradiation conditions through a wide dose rate range, from 0.01 Gy/s up to about $10^5$ Gy/s, at various oxygen concentrations. The preliminary results indicate a correlation between the temporal structure of the pulses and a significant reduction in the production of reactive oxygen species (ROS) at different dose rates.

**Keywords:** water radiolysis, FLASH effect, UHDR, oxygenation, Geant4-DNA

## 1. Introduction

FLASH radiotherapy ("FLASH") is characterized by a significant increase in normal tissue tolerance during irradiation at an ultra-high dose rate (UHDR) while keeping efficient control in tumour tissue. This non-conventional technique requires delivery dose rates larger than 40 Gy/s for a given radiation dose and demonstrates greater normal tissue sparing compared to conventional irradiation (≤ 0.3 Gy/s, "CONV") [1][2][3]. However, while these observations have been confirmed *in vivo* [1], the connection between the high-density energy deposition during short-duration physical events (ionizations or excitations) and biological responses is still not fully understood. In addition to considering biological processes such as inflammation or redox biology, recent studies investigating oxygenation dependence suggest a crucial role of the chemical phase in the observed *in vivo* endpoints [1]. Since about 70% of the mass in cells and tissues is water, the role of radiolysis in water is of great interest for such radiobiological applications.

The Monte Carlo technique is a valuable approach for describing the kinetics of species diffusion and their reactions under the influence of high-density interacting radiation tracks in water radiolysis. This scenario is particularly relevant for FLASH irradiation, as it allows for the study of the chemical effects induced by a large number of localized tracks present at the same time or following a time sequence in the irradiated volume. Several Monte Carlo-based software packages have been developed to date, including IONLYS-IRT [4], TOPAS-nbio [5], TRAX-CHEM [6], Geant4-DNA [7-11] and others. These codes enable the description of water radiolysis over time through a variety of approaches.

The Geant4-DNA [7-11] extension of Geant4 [12-14] can model both the physical and chemical stages of water radiolysis. The chemical stage is currently simulated using either the particle-based step-by-step (SBS) Brownian dynamics [15,16] or the Independent Reaction Time (IRT) [17-21] models to resolve the heterogeneous distributions up to 1 μs [11]. Although the level of detail is limited in these microscopic particle-based models (i.e., spherical particles representing the chemical species in a continuum medium representation), the calculation time is still the main critical point when simulations deal with a large number of species or for long-time scales [11, 22] as encountered for the FLASH effect. Another approach has been recently implemented in Geant4-DNA using the compartment-based representation [22]. By gathering the same species as one group placed in a small enough volume (so-called "compartment" or "voxel") where the distribution of species is assumed to be homogeneous, the evolution of species is described by their concentration through the reaction-diffusion master equation (RDME). Under this condition, this mesoscopic model allows simulations of a high number of species. The chemical SBS and RDME models implemented in Geant4-DNA have been combined within the new Geant4-DNA example [22] named "UHDR" available from version 11.2 of Geant4. This example enables users to simulate the time-dependent radiation-chemical yields (G-values) of chemical species across multiple time scales, starting from 1 picosecond (the initiation of the chemical stage) and extending to several minutes, while considering acid-base equilibria under various conditions of oxygen concentration and dose-rate. Under such oxygenation and dose rate conditions, the simulation can predict time-dependent yields of free radical species such as hydroxyl radicals ($^{\bullet}OH$), hydrated electrons ($e^-_{aq}$), and reactive oxygen species (ROS) produced in dissolved oxygen. Note that, in this work, we define ROS as superoxide anion radical ($O_2^{\bullet-}$), hydroperoxyl radical ($HO_2^{\bullet}$), and hydrogen peroxide ($H_2O_2$).

One current limitation of the "UHDR" example is that particle beams are simulated as a single instantaneous pulse (also called "infinite" pulse). This infinite pulse means that all incident particles are emitted simultaneously, with no pulse duration. Such irradiation conditions differ from reality, where the absorbed dose is the same for CONV and FLASH, and it is delivered over a long train of individual pulses, for a beam duration of up to a few minutes. In this study, we introduce a new combined model between the IRT and the RDME models to simulate water radiolysis under single pulse irradiation and taking into account the time duration of the single pulse, for different absorbed doses of 0.01, 0.1, 1, and 2 Gy and in an oxygenated water volume. Such single pulses have been sampled from a real raw signal of a ultra-high dose rate (UHDR) pulse measured by an eRT6/Oriatron linac accelerator (PMB-Alcen, France) described in [23].

In the next section, we present the principles of this combined model, including the physics and chemistry simulation scheme. The simulations of water radiolysis are conducted at various oxygen concentrations and at a pH of 5.5 over long timescales. This pH value was chosen to reflect the measured pH resulting from $CO_2$ absorption in water during the purification process [24].

## 2. Materials and methods

### 2.1. Simulation scheme

#### 2.1.1. Irradiation

The simulation is performed using the Geant4-DNA "UHDR" example with the geometry being a cubic water volume of 3.2 x 3.2 x 3.2 $\mu m^3$ (Fig. 1) or 1.6 x 1.6 x 1.6 $\mu m^3$ size, irradiated with electrons in two different ways: instantaneous single pulse irradiation (all primary particles are shot together at the same time, "infinite" pulse) or irradiation considering a time sequence (electrons are shot individually at different times during the single pulse). Since energy deposition is proportional to the size of the volume for a given absorbed dose, the choice of the simulation volume is a compromise between a large enough number of chemical species and an achievable computational time. The periodic boundary condition (PBC) is used to simulate the behavior of secondary electrons during physical stage. When an electron exits an edge of a cubic volume, it re-enters from the opposite edge. The PBC [25] helps reduce the edge effects in dose calculations for micrometer-sized volumes. This approach mimics the behavior of a macroscopic volume without requiring a large computational resource. In this work, PBC is only applied for secondary (and subsequent) electrons produced during the physical stage.

For each simulation, a 1 MeV electron source is randomly generated on one of the faces of the volume with normal incidence. The physics list "G4EmDNAPhysics_option2" is used to model the physical stage. Once the physical stage concludes at the considered absorbed dose per pulse, the physico-chemical stage dissociates excited and ionized water molecules along electron tracks. The physico-chemical stage is simulated using the "G4ChemDissociationChannels_option1" class [26], available in the "G4EmDNAChemistry_option3" [26] chemistry constructor. After the physico-chemical stage, early reactant species (or primary chemical species) such as $e^-_{aq}$, $H_2$, $H^\bullet$, $^\bullet OH$, $OH^-$, $H_3O^+$, $O(3P)$ are created during the physico-chemical stage form their heterogeneous distributions along energy deposit regions of track structure (so-called "spurs").

### 2.1.2. Time structure of a single pulse

The public version of Geant4 includes the Geant4-DNA "UHDR" example (Geant4 11.2). This example can generate a 1 MeV electron source by default. Primary electrons are emitted simultaneously until the sum of all energy deposits in the volume reaches a specified absorbed dose. We refer to this as an "instantaneous pulse" or "infinite pulse" (see Figure 1 – left), where all primary chemical species induced by the electron irradiation are produced simultaneously within 1 ps. This approach can be challenging to compare with experimental data, where the time sequence of successive electrons in a pulse time structure may significantly impact observations in the chemistry phase or biological endpoints.

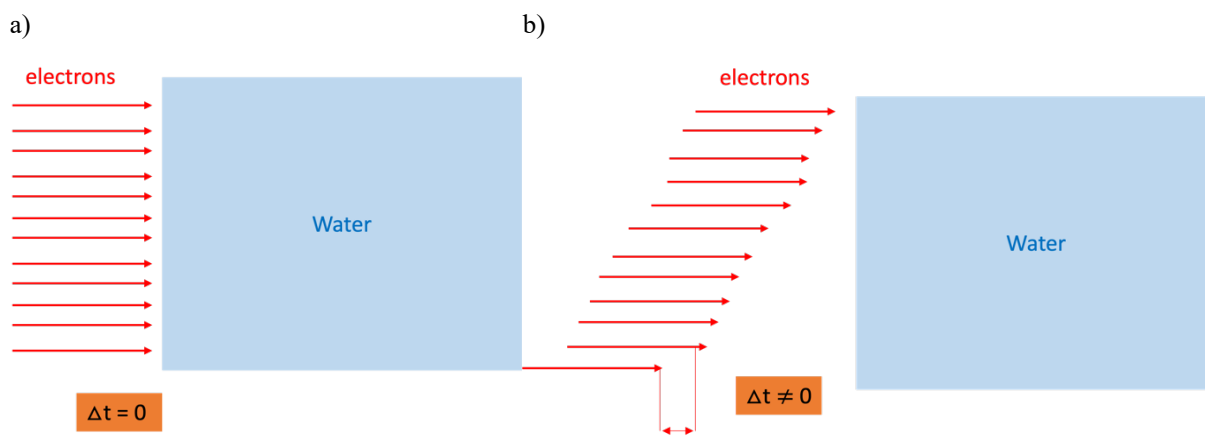

*Figure 1: 1 MeV electron beam irradiation. Simulation volume: water cube (in blue). a): Infinite (= instantaneous) single-pulse irradiation: all incident electrons are shot simultaneously as a parallel beam onto the volume until the total absorbed dose is reached. b): « Time-structured » single-pulse irradiation. The sequence of irradiation can be changed by the user. Δt is the delay time between two successive electrons within a beam irradiation.*

In Geant4-DNA, a time structure for single-pulse irradiation can be activated by implementing a delay time, $\Delta t$, between the relative time associated with the primary particles; this $\Delta t$ can be sampled from a beamline's raw signal of a measured pulse obtained from particle accelerators. This delay time is then propagated to the secondary electrons, as well as to the corresponding solvated electrons and dissociated water fragments, such as ionized and excited water molecules, which are induced by the primary and secondary particles. These highly energized water fragments and the solvated electrons then remain inactive throughout the irradiation duration. The primary particles continue to irradiate the volume until their deposited energies reach the desired absorbed dose, at which point the physical stage ends.

In this study, we investigated different doses, 0.01 Gy, 0.1 Gy, 1 Gy and 2 Gy, with pulse durations ranging from instantaneous (infinite) pulse up to 1 s. For short pulse durations such as 1.4 µs, 2.4 µs, and 3.5 µs, the delay time can be sampled from measured beamline raw signals. In this work, the raw signal from the high dose-per-pulse linear accelerator Oriatron/eRT6 [23] is used to generate a series of delay times (see Figure 2). The objective is to account for the time structure characteristics, including the pulse duration, the pulse repetition frequency, and the pulse amplitude for the different beam irradiation conditions. The longer pulse widths such as 1 ms, 10 ms, 100 ms, 500 ms, or 1 second are considered as continuous beam structures using a uniform distribution to sample delay times until the last primary particle and its secondary electrons deposit their energies; such moment is referred to as the "final deadline". By dividing the obtained dose by this "final deadline" time, we obtain a dose rate. This allowed us to select a dose rate ranging from approximately a dose-per-pulse of about $10^5$ Gy/s for the higher dose rate down to 0.01 Gy/s for the lower dose rate.

Once the physical stage of all primary particles is complete, the chemistry stage begins. A time scheduler matches the delay times with the virtual time of the simulation, successively activating the corresponding solvated electrons and fragmented water molecules. Upon activation, the physico-chemical dissociation processes convert the water fragments into primary chemical species, integrating these reactive species into the system. These species then are synchronized with previous chemical species. All these reactive species can then diffuse and react with each other or with previously formed primary chemical species. This integration continues until the last delay time, which corresponds to the final deadline matched with the virtual time, ensuring that all primary chemical species are fully integrated into the system. This final deadline is then referred to as the "integration duration" or pulse duration.

*a)* 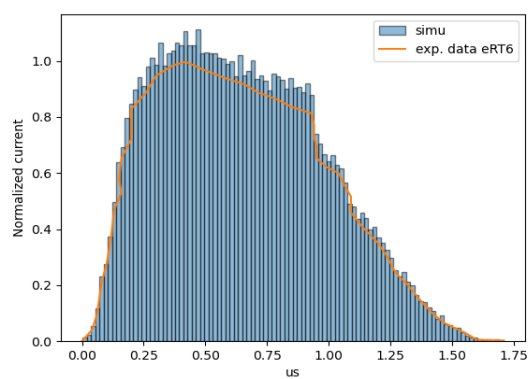 *b)* 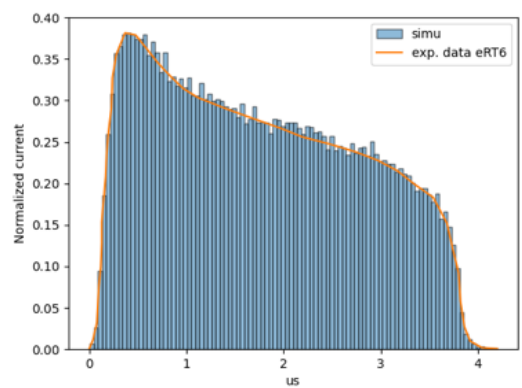

*c)*

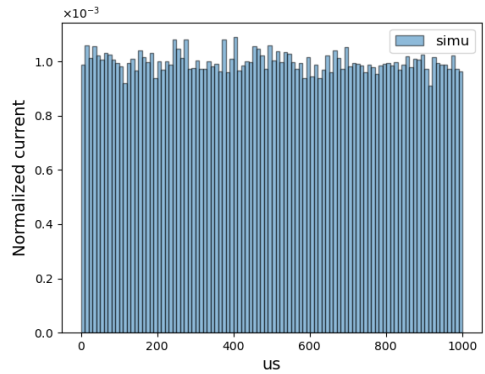

*Figure 2: The pulse raw beam signal as a function of time obtained from the high dose-per-pulse linear accelerator Oriatron eRT6 measured (orange line) and sampled delay time used in Geant4-DNA simulations (blue histograms). Examples of four types of single pulses with different durations are shown: a) 1.4 μs, b) 3.5 μs, c) 1 millisecond. The 1.4 μs and 3.5 μs duration pulse raw signals are based on [23].*

*2.1.3. Time structure of multi-pulse irradiation*

Since the model allows for the extension of pulse duration up to 1 second, a multi-pulse temporal structure can be obtained by applying a pulse repetition frequency, including the temporal pulse width and the dose-per-pulse. To characterize the time structure of pulse repetition frequency, Figure 3 provides a typical example defining the Duration of the Train (DT) and the Duration Inter-Train (DIT) [27]. This illustration shows that UHDR typically features high dose-per-pulse and short DIT intervals, facilitated by high repetition frequencies, whereas CONV is characterized by longer DIT intervals with a series of lower dose-per-pulse. These typical beam time structures consist of irradiation output signals with DIT ranging from 1 to 10 microseconds, delivered at repetition rates between 10 and 1000 Hz [3].

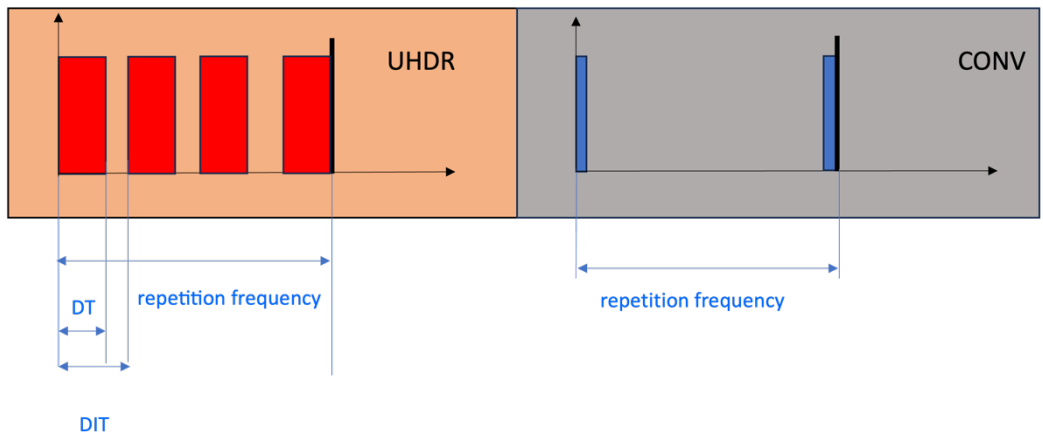

*Figure 3: Examples of a typical UHDR time structure irradiation (one train of 4 pulses) and CONV (one train of 2 pulses) irradiation during one repetition frequency (limited by the vertical back line). The rectangles (red or blue) represent an individual beam pulse.*

These features can be partially reproduced using the Geant4-DNA "UHDR" example using a dedicated user interface. Table 1 presents the simulated beam parameters of multi-pulse time structures for both CONV (0.02 Gy/s mean dose rate) and UHDR (up to 2 Gy/s mean dose rate) modes, including the DIT, the pulse number, the mean dose rates and the average incident particle number needed to reach the selected absorbed dose in corresponding volume. Note that, in this study, we define UHDR as a dose rate of 1 Gy/s or higher and CONV as up to 0.1 Gy/s. The pulse repetition frequencies (PRF) result in pulse intervals (DIT) of approximately 10 milliseconds for UHDR and 100 milliseconds for CONV [28] (see Section 4). Delay times were sampled throughout the beam time structures during irradiation. The sampling is similar to the one of a single pulse, except that particles are discretely integrated over the pulse duration (DT). This pulse shape is repeated to simulate a long train of pulses (up to 50 pulses) during 500 ms of beam irradiation duration. To compare the impact of beam structure, a continuously single pulse of 500 ms (see Section 2.1.2) is included in this comparison using the same dose.

Figure 4 (top) shows an example of a simulated train of pulses with 50 pulses and a DIT of 10 ms (PRF = 100 Hz), compared with a sampled single pulse of 500 ms. While each pulse of the pulse trains is sampled using the 1.4 µs raw signal pulse shape [23] (Figure 4, bottom), the single pulse of 500 ms is sampled using a uniform distribution. The lower repetition frequency of DIT = 100 ms (PRF = 10 Hz) is shown in Appendix: Figure 1.

| Mode | DIT (ms) | Pulse number | Mean dose rate (Gy/s) | Volume (µm³) | Average incident electrons |
|---|---|---|---|---|---|
| **CONV** | 100 | 5 | 0.02 | 6.4 x 6.4 x 6.4 | 19 |
|  | Single pulse | 1 | 0.02 |  |  |
|  | 10 | 50 | 0.2 |  | 154 |
| **UHDR** | 10 | 50 | 1 | 3.2 x 3.2 x 3.2 | 161 |
|  | Single pulse | 1 | 2 |  |  |
|  | 10 | 50 | 2 |  | 328 |
|  | 100 | 5 | 2 |  |  |
|  | 10 | 50 | 4 |  | 715 |
|  | Single pulse | 1 | 4 |  |  |

*Table 1: Simulated beam parameters for CONV and UHDR modes for an irradiation duration of 500 ms.*

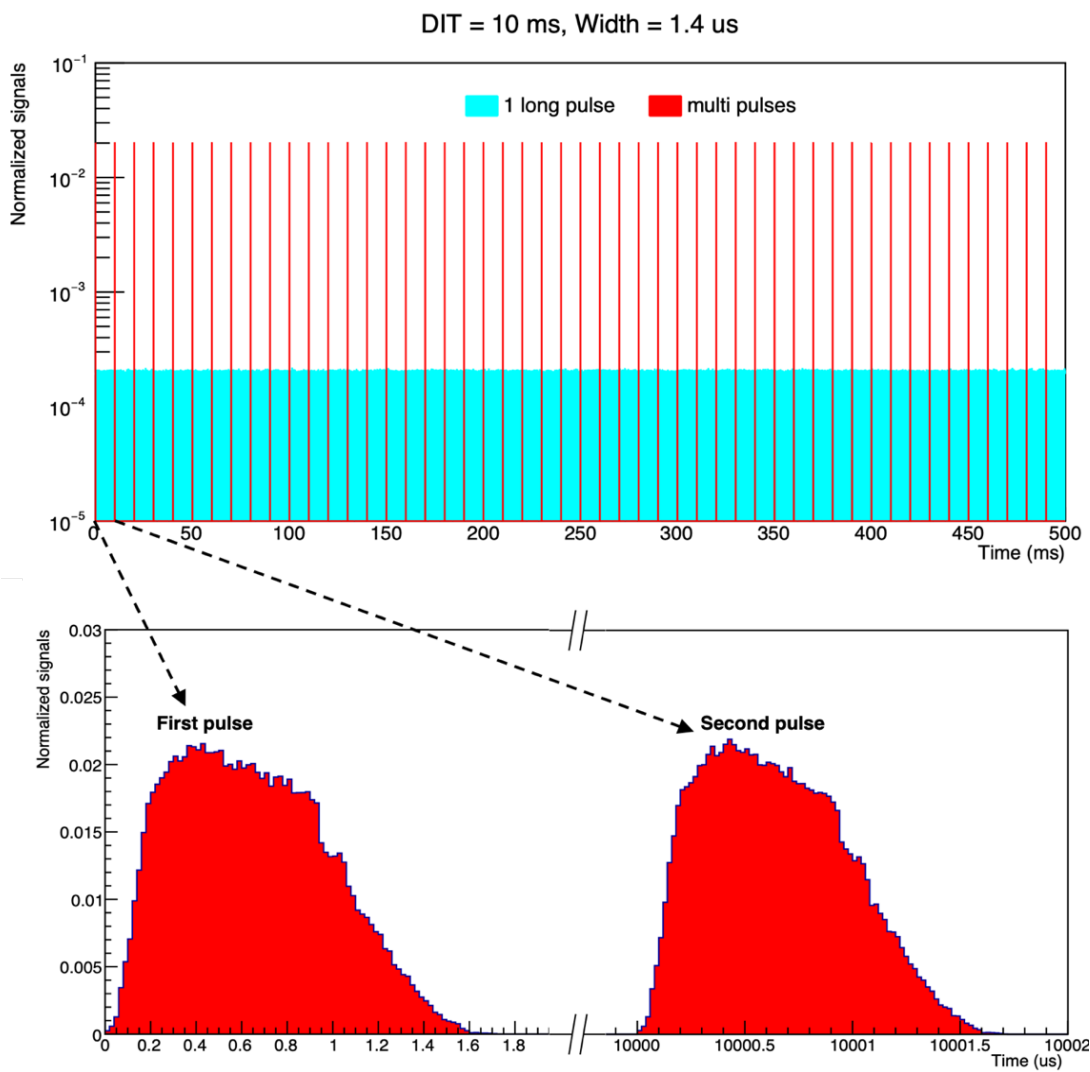

*Figure 4: Top: an example of beam time structure used in the simulation of 50 pulses (red lines) with a DIT of 10 ms and a long single pulse (blue rectangle) over 500 milliseconds. Bottom: Pulse shape of the first two pulses.*

*2.2. Update of the synchronized independent reaction time approach*

During the integration duration, we applied the particle-based model using the IRT approach until the "final deadline" when the last primary chemical species was integrated into the system. Once the simulation reached 5 nanoseconds after the final deadline, we stopped the particle-based IRT simulation and transitioned to a uniform 3D Cartesian mesh to benefit from the efficiency of the mesoscopic model [22].

To achieve long beam irradiation up to 1 second using the IRT approach, this work introduced a new optimization of the synchronized IRT [20] (so-called IRT-syn model). The reaction list is now computed at the beginning of the radiolysis simulation and is not resampled at each time step as was done in the previous version, except when the involved species interact with the medium (where medium interaction refers to interactions between species and scavengers or volume limits) or when new primary chemical species are integrated during the pulse duration. Each time step concludes with diffusion, synchronized with the time step. Figure 5 illustrates the flow process. The event table is created to gather the reaction list using the IRT approach and medium interaction. Each event corresponds with a reaction. The next event is evaluated to check if its time step corresponds to the end time of the simulation. If it does not, the next event is processed, and all species are synchronized through diffusion with the time step. During this process, the species are checked to determine if they interact with the medium (for example, $e^-_{aq}$ reacts with oxygen scavenger [30] or reaches the volume borders).

- If yes, a new event table is recomputed based on the new positions of all species;
- If no, the next reaction of the current event table is processed.

The next section will present the algorithms for volume boundary interactions.

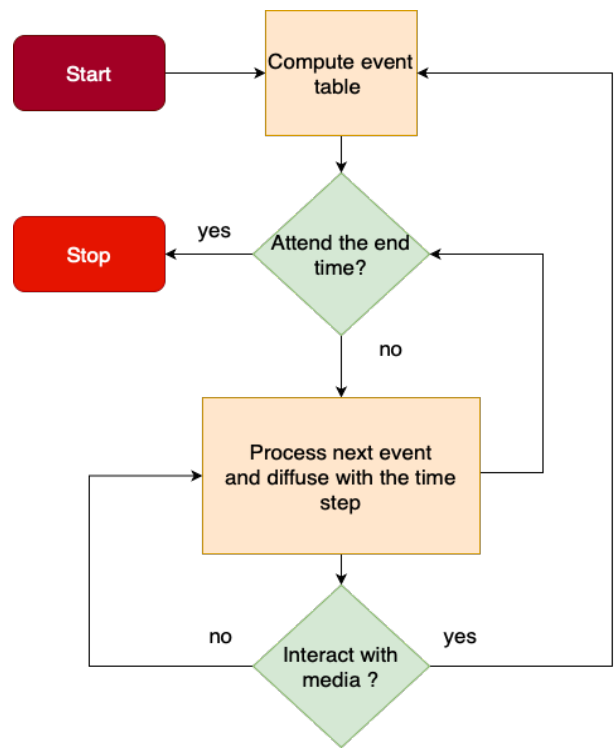

*Figure 5: The flow process of the stepping algorithm implemented in Geant4-DNA.*

2.3. Bounded volume

During the chemical stage, we consider that molecules diffuse and react in a bounded volume (that is, the diffusion is limited by geometrical boundaries) which is also the irradiated water box volume of the physical stage. The bouncing of chemical molecules on the limited volume border to confine the chemical molecules in the considered volume is applied for both microscopic and mesoscopic sub-stages, depicting a closed system of experimental test cells as typically used for *in vitro* measurements. The setup is shown in Figure 6, for a 3.2 x 3.2 x 3.2 µm³ simulation setup.

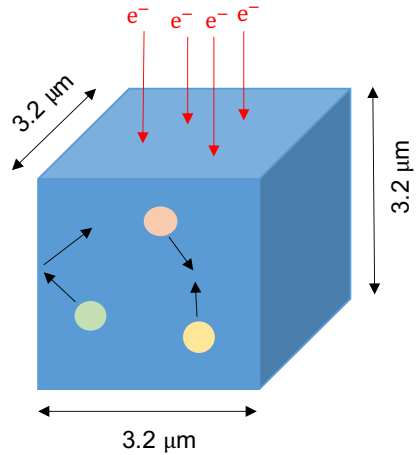

*Figure 6: Simulation of the water box and its irradiation by electrons. The spheres inside the water box represent chemical species that interact with each other. The spheres will bounce back into the box when encountering a boundary of the box.*

Initially introduced by Karamitros et al. [19], the analytical reflective box model was developed based on an algorithm that allows for the sampling of the next position of a Brownian particle located in a reflective box after a time step. Details of the algorithm and the comparison between the sampling algorithm and the analytical solution can be found elsewhere in [19]. In this work, we use this algorithm to sample the next position of chemical molecules at the boundary of the volume after a time step $\Delta t$, and the molecules are reflected if they reach the boundary. This time step to boundaries is computed by:

$$R = erfc(y)$$

This formula was initially introduced by Karamitros et al. [16] in Geant4-DNA to compute time and space intervals to reach a volume boundary where $y = \frac{d}{2\sqrt{D\Delta t}}$, $erfc(x) = 1 - erf(x)$, $D$ is the diffusion coefficient, $R$ is a uniform random number in [0,1], $d$ is the closest distance between species and the boundary of the volume and $\Delta t$ is the time step. The sampling of this diffusion event is computed at each time step for all species to prevent species from escaping the volume.

### 2.4. Oxygenation

The implementation of oxygenation in water was presented elsewhere in [30]. The resulting oxygen depletion for the different dose rates is recorded over time during the chemical stage. We tested partially oxygenated water concentrations of 4% and 21% in the simulation volume. While 4% represents hypoxia conditions, 21% of oxygen corresponds to atmospheric conditions.

### 2.5. List of chemical reactions

The reaction scheme is classified into three categories according to their time scales:

### 2.5.1. Primary reactions

Primary reactions happen from a few picoseconds to a few microseconds, involving the various initial physico-chemical products ($e^-_{aq}$, H•, •OH, OH$^-$, H$_3$O$^+$). Due to high concentrations of these products in heterogeneous spurs and their high-rate constants, these reactions (Table 2) nearly dominate the heterogeneous periods.

| Reactions | k (L.mol$^{-1}$.s$^{-1}$) |
|---|---|
| H$^{\bullet}$ + e$^{-}_{aq}$ → H$_2$ + OH$^{-}$ | 2.50 x 10$^{10}$ |
| H$^{\bullet}$ + H$^{\bullet}$ → H$_2$ | 5.03 x 10$^9$ |
| H$^{\bullet}$ + $^{\bullet}$OH → H$_2$O | 1.55 x 10$^{10}$ |
| $^{\bullet}$OH + e$^{-}_{aq}$ → OH$^{-}$ | 2.95 x 10$^{10}$ |
| $^{\bullet}$OH + $^{\bullet}$OH → H$_2$O$_2$ | 5.50 x 10$^9$ |
| e$^{-}_{aq}$ + H$_2$O$_2$ → OH$^{-}$ + $^{\bullet}$OH | 1.10 x 10$^{10}$ |
| H$_3$O$^{+}$ + OH$^{-}$ → H$_2$O | 1.13 x 10$^{11}$ |
| e$^{-}_{aq}$ + H$_3$O$^{+}$ → H$^{\bullet}$ | 2.11 x 10$^{10}$ |
| e$^{-}_{aq}$ + e$^{-}_{aq}$ → H$_2$ + OH$^{-}$ | 6.36 x 10$^9$ |

Table 2: Primary reactions taking place from a few picoseconds to a few microseconds, involving the various initial physico-chemical products (e$^{-}_{aq}$, H$^{\bullet}$, $^{\bullet}$OH, OH$^{-}$, H$_3$O$^{+}$) [29, 26].

### 2.5.2. Secondary reactions and reactions with oxygen

Secondary reactions involve the participation of primary reaction products and reactions with the solvent medium. In oxygenated water, the following secondary reactions are associated with the interaction of chemical products ($^{\bullet}$OH, e$^{-}_{aq}$, H$^{\bullet}$) that have "escaped" from the heterogeneous spurs, with the solvent medium, such as oxygen in reactions (1) and (2). These interactions occur over a range of timescales, which can take place from a few hundred nanoseconds to minutes, depending not only on their reaction constants but also on their concentrations. This range of timescales can include the recombination of OH$^{-}$, and H$_3$O$^{+}$ ions and may initiate acid-base reactions with the water medium. The reactions are shown in Table 3.

$$e^{-}_{aq} + O_2 \rightarrow O_2^{\bullet -} \qquad (1)$$

$$H^{\bullet} + O_2 \rightarrow HO_2^{\bullet} \qquad (2)$$

| Reaction | Reaction rate (L.mol$^{-1}$.s$^{-1}$) | Reaction | Reaction rate (L.mol$^{-1}$.s$^{-1}$) |
|---|---|---|---|
| H$^{\bullet}$ + O$^{\bullet -}$ → OH$^{-}$ | 2.00 x 10$^{10}$ | H$^{\bullet}$ + H$_2$O$_2$ → $^{\bullet}$OH + H$_2$O | 3.50 x 10$^7$ |
| $^{\bullet}$OH + HO$_2^{-}$ → HO$_2^{\bullet}$ + OH$^{-}$ | 8.32 x 10$^9$ | H$^{\bullet}$ + O$_2$ → HO$_2^{\bullet}$ | 2.10 x 10$^{10}$ |
| $^{\bullet}$OH + O$^{\bullet -}$ → HO$_2^{-}$ | 1.00 x 10$^9$ | H$^{\bullet}$ + O$_2^{\bullet -}$ → HO$_2^{-}$ | 1.00 x 10$^{10}$ |
| $^{\bullet}$OH + HO$_2^{\bullet}$ → O$_2$ + H$_2$O | 7.90 x 10$^9$ | H$_2$ + O$^{\bullet -}$ → H$^{\bullet}$ + OH$^{-}$ | 1.21 x 10$^8$ |
| $^{\bullet}$OH + H$_2$O$_2$ → HO$_2^{\bullet}$ + H$_2$O | 2.88 x 10$^7$ | O$_2$ + O$^{\bullet -}$ → O$_3^{-}$ | 3.70 x 10$^9$ |
| $^{\bullet}$OH + H$_2$ → H$^{\bullet}$ + H$_2$O | 3.28 x 10$^7$ | HO$_2^{\bullet}$ + HO$_2^{\bullet}$ → H$_2$O$_2$ + O$_2$ | 9.80 x 10$^5$ |
| •OH + O₂•⁻ → O₂ + OH⁻ | 1.07 x 10¹⁰ | HO₂• + O₂•⁻ → HO₂⁻ + O₂ | 9.70 x 10⁷ |
| •OH + O₃⁻ → O₂•⁻ + HO₂• | 8.50 x 10⁹ | e⁻_aq + O•⁻ + H₂O → 2 OH⁻ | 2.31 x 10¹⁰ |
| e⁻_aq + HO₂• → HO₂⁻ | 1.29 x 10¹⁰ | e⁻_aq + O₂•⁻ + H₂O → H₂O₂ + 2 OH⁻ | 1.29 x 10¹⁰ |
| e⁻_aq + O₂ → O₂•⁻ | 1.74 x 10¹⁰ | e⁻_aq + HO₂⁻ → O⁻ + OH⁻ | 3.51 x 10⁹ |
| H₂O₂ + O•⁻ → HO₂• + OH⁻ | 5.55 x 10⁸ | O•⁻ + O•⁻ + 2 H₂O → H₂O₂ + 2 OH⁻ | 1.00 x 10⁸ |
| H• + HO₂• → H₂O₂ | 1.00 x 10¹⁰ | O•⁻ + O₃⁻ → 2 O₂•⁻ | 7.0 x 10⁸ |
| O•⁻ + HO₂⁻ → O₂•⁻ + OH⁻ | 3.50 x 10⁸ | O•⁻ + O₂•⁻ + H₂O → O₂ + 2 OH⁻ | 6.00 x 10⁸ |

Table 3: Secondary reactions taking place from a few microseconds to minutes, involving the various secondary products (HO₂•, O₂•⁻, H₂O₂...) [29,31].

### 2.5.3. Acid-base reactions

The products of primary and secondary reactions can participate in equilibrium reactions which are associated with pKa (Table 4). These processes can start from a few microseconds to a very long time when steady states are reached.

| # | Equilibrium reactions | pKa | $K_{\#}$ |
|---|---|---|---|
| 1 | 2 H₂O ↔ OH⁻ + H₃O⁺ | 13.99 | 9.85 x 10⁻¹⁵ M |
| 2 | H₂O₂ + H₂O ↔ HO₂⁻ + H₃O⁺ | 11.65 | 1.65 x 10⁻¹² M |
| 3 | •OH + H₂O ↔ O•⁻ + H₃O⁺ | 11.9 | 1.26 x 10⁻¹² M |
| 4 | HO₂• + H₂O ↔ O₂•⁻ + H₃O⁺ | 4.57 | 1.49 x 10⁻⁵ M |
| 5 | H• + H₂O ↔ e⁻_aq + H₃O⁺ | 9.77 | 2.81 x 10⁻¹ M |

Table 4: pKa of the primary equilibria and water at 25°C. $K_{\#}$ represents the first equilibrium constant, $K_1$, for these reactions [32].

Based on the H₃O⁺ and OH⁻ ion concentrations determined by the pH, acid-base reactions associated with these pKa are simulated. These ion concentrations are treated as constants throughout the simulation. Other equilibrium processes, based on these concentrations, are shown in Table 5. The pH is fixed at 5.5 in these simulations [24].

| # | Acid-Base reactions | Rate coefficients and corresponding references | |
|---|---|---|---|
| 1 | HO₂• → H₃O⁺ + O₂•⁻ | $k_{-1} * K_4$ | 7.58 x 10⁵ s⁻¹ |
| −1 | H₃O⁺ + O₂•⁻ → HO₂• | [31] | 4.78 x 10¹⁰ M⁻¹s⁻¹ |
| 2 | H• → e⁻_aq + H₃O⁺ | $k_{-2} * K_5$ | 6.32 s⁻¹ |
| −2 | e⁻_aq + H₃O⁺ → H• | [31] | 2.25 x 10¹⁰ M⁻¹s⁻¹ |
| 3 | $e^-_{aq} + H_2O \rightarrow H^{\bullet} + OH^-$ | $k_{-3} * K_1 / (K_5 * [H_2O])$ | $1.57 \times 10^1$ M$^{-1}$s$^{-1}$ |
| $-3$ | $H^{\bullet} + OH^- \rightarrow H_2O + e^-_{aq}$ | [Elliot, 1994] | $2.49 \times 10^7$ M$^{-1}$s$^{-1}$ |
| 4 | $O_2^{\bullet-} + H_2O \rightarrow HO_2^{\bullet} + OH^-$ | $k_{-4} * K_1 / (K_4 * [H_2O])$ | 0.15 M$^{-1}$s$^{-1}$ |
| $-4$ | $HO_2^{\bullet} + OH^- \rightarrow O_2^{\bullet-} + H_2O$ | [31] | $1.27 \times 10^{10}$ M$^{-1}$s$^{-1}$ |
| 5 | $HO_2^- + H_2O \rightarrow H_2O_2 + OH^-$ | $k_{-5} * K_1 / (K_2 * [H_2O])$ | $1.36 \times 10^6$ M$^{-1}$s$^{-1}$ |
| $-5$ | $H_2O_2 + OH^- \rightarrow HO_2^- + H_2O$ | [31] | $1.27 \times 10^{10}$ M$^{-1}$s$^{-1}$ |
| 6 | $O^{\bullet-} + H_2O \rightarrow {}^{\bullet}OH + OH^-$ | $k_{-6} * K_1 / (K_3 * [H_2O])$ | 1.8e6 M$^{-1}$s$^{-1}$ |
| $-6$ | ${}^{\bullet}OH + OH^- \rightarrow O^{\bullet-} + H_2O$ | [31] | $1.27 \times 10^{10}$ M$^{-1}$s$^{-1}$ |
| 7 | $H_2O_2 \rightarrow H_3O^+ + HO_2^-$ | $k_{-7} * K_2$ | $7.86 \times 10^{-2}$ s$^{-1}$ |
| $-7$ | $HO_2^- + H_3O^+ \rightarrow H_2O_2$ | [31] | $4.78 \times 10^{10}$ M$^{-1}$s$^{-1}$ |
| 8 | ${}^{\bullet}OH \rightarrow O^{\bullet-} + H_3O^+$ | $k_{-8} * K_3$ | 0.0602 s$^{-1}$ |
| $-8$ | $O^{\bullet-} + H_3O^+ \rightarrow {}^{\bullet}OH$ | [31] | $9.56 \times 10^{10}$ M$^{-1}$s$^{-1}$ |

*Table 5: Acid-base reactions associated with pKa at 25 °C. [H₂O] = 55.3 M. K_# in Table 4 is used to define reaction rate constants in these processes. k₋₁ and k₁ represent opposite directions of $H_3O^+ + O_2^{\bullet-} \rightarrow HO_2^{\bullet}$ and $HO_2^{\bullet} \rightarrow H_3O^+ + O_2^{\bullet-}$, respectively, and so on for the other reactions [31].*

## 3. Results

### 3.1 Benchmarking of IRT implementation

The IRT implementation in a bounded volume is compared with the analytical solution of the simplest second-order reaction of identical reactants ${}^{\bullet}OH + {}^{\bullet}OH \rightarrow H_2O_2$. The differential rate law is proportional to the square of the concentration of the reactant:

$$\frac{1}{2}\frac{d[A]}{dt} = -k[A]^2$$

The integrated rate law describes the concentration of the reactant at a given time:

$$\frac{1}{[A_t]} = \frac{1}{[A_0]} + 2kt$$

where $A_t$ is the reactant concentration at time $t$, $A_0$ is the initial reactant concentration and $k$ is the reaction rate.

The IRT simulation homogenously generated 200 species in a 1.6 x 1.6 x 1.6 μm³ bounded volume for the reaction ${}^{\bullet}OH + {}^{\bullet}OH \rightarrow H_2O_2$, (with) $k = 5.50 \times 10^9 \; (L.mol^{-1}.s^{-1})$ (see Table 2). As can be seen in Figure 7, under this reflective model, a good agreement is observed between our implemented version of IRT and the analytical model. More benchmarking tests can be found elsewhere in [19].

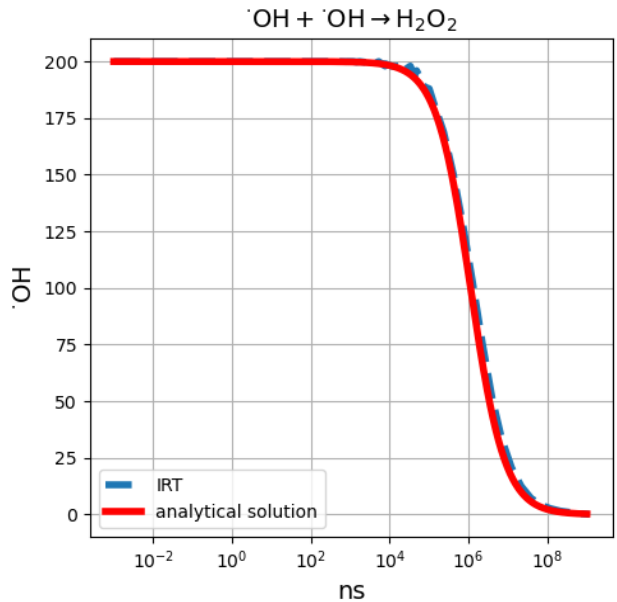

*Figure 7: Comparison between the IRT implementation and the analytical model in a 1.6 x 1.6 x 1.6 μm³ bounded volume.*

While the first benchmarking is done based on initially homogeneous distribution of •OH species to compare with the analytical solution, the second test verifies the chemical species evolution from high-concentration species of early radiation-induced spurs to homogeneous kinetics in bulk solution. The simulations were performed using the "UHDR" example capable of generating a 1 MeV electron source. These electrons are shot randomly onto one of the faces of a 3.2 x 3.2 x 3.2 μm³ cubic volume (see Figure 1-a) of water to accumulate an absorbed dose of 0.1 Gy. When this desired dose is reached, all primary chemical species induced by the irradiation are produced simultaneously at 1 ps. The reaction list only applied the •OH + •OH → $H_2O_2$ reaction. The IRT-only implementation is compared with the SBS model combined with the mesoscopic model (RDME) (so-called "SBS-RDME") [22], and with the IRT model combined with the mesoscopic model (so-called "IRT-RDME"). Note that the RDME is activated from 5 ns for both combination models. Figure 8 shows that the IRT implementation can successfully reproduce the •OH recombination from the initial heterogeneous distribution to the homogeneous states up to 1 second. This benchmarking also demonstrates the capability of IRT-syn to provide sufficient accurate spatio-temporal information of the species for the RDME (or mesoscopic) model [22].

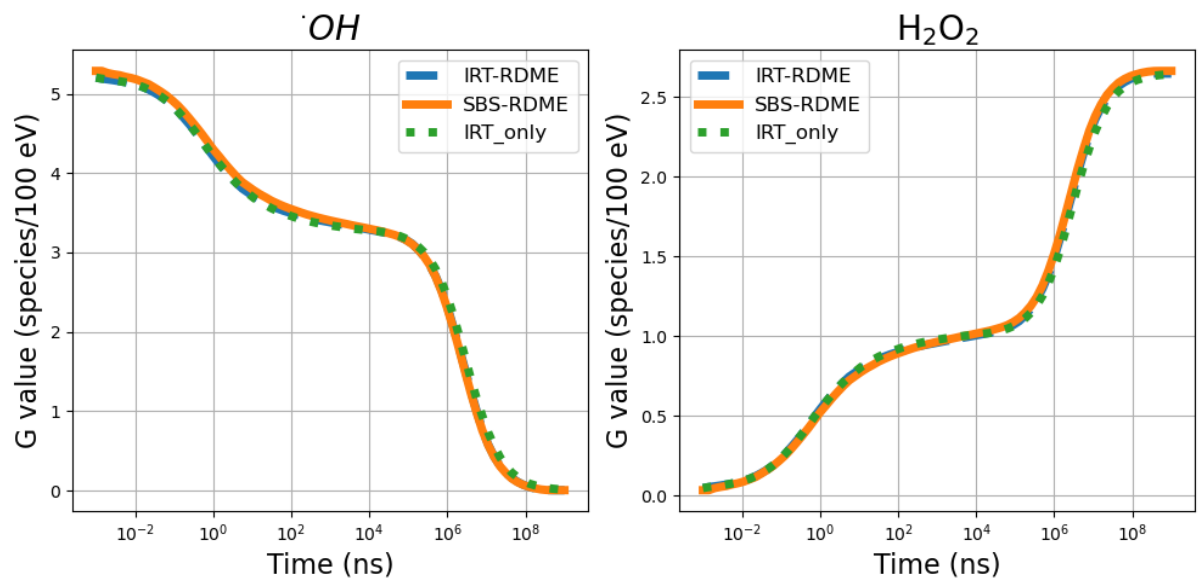

*Figure 8: Comparison between IRT-only implementation, SBS-meso and IRT-meso, from 1 ps up to 1 s post irradiation, for •OH (left) and $H_2O_2$ (right) species.*

*3.2. Species evolution considering the time structure of the pulse*

Figure 9 shows the comparison of the time-dependent G-values of •OH, $e^-_{aq}$, $O_2^{•-}$, and $H_2O_2$ up to about 15 minutes after irradiation with a 1 MeV electron beam into a cubic water volume of 3.2 x 3.2 x 3.2 μm³, with 21% oxygenation, resulting in a total absorbed dose of 0.1 Gy. The comparisons were made for an instantaneous pulse and different pulse durations: 1.4 μs, 2.4 μs, 3.5 μs, 1 millisecond (ms), 10 ms, 100 ms, 500 ms, and 1 second (s), corresponding to dose rates ranging from 0.1 Gy/s to $7.14 \times 10^4$ Gy/s for a single pulse. The simulation is repeated 100 times to ensure that the statistical uncertainty remains below 5%. The time structure in one pulse is necessary for the short pulse durations only (1.4 μs, 2.4 μs and 3.5 μs). For the longer pulse durations of 1 ms, 10 ms, 100 ms, and 1 second, because of the short drop of beam stops, the beam current is kept constant long enough to be considered as a uniform distribution (see Figure 2).

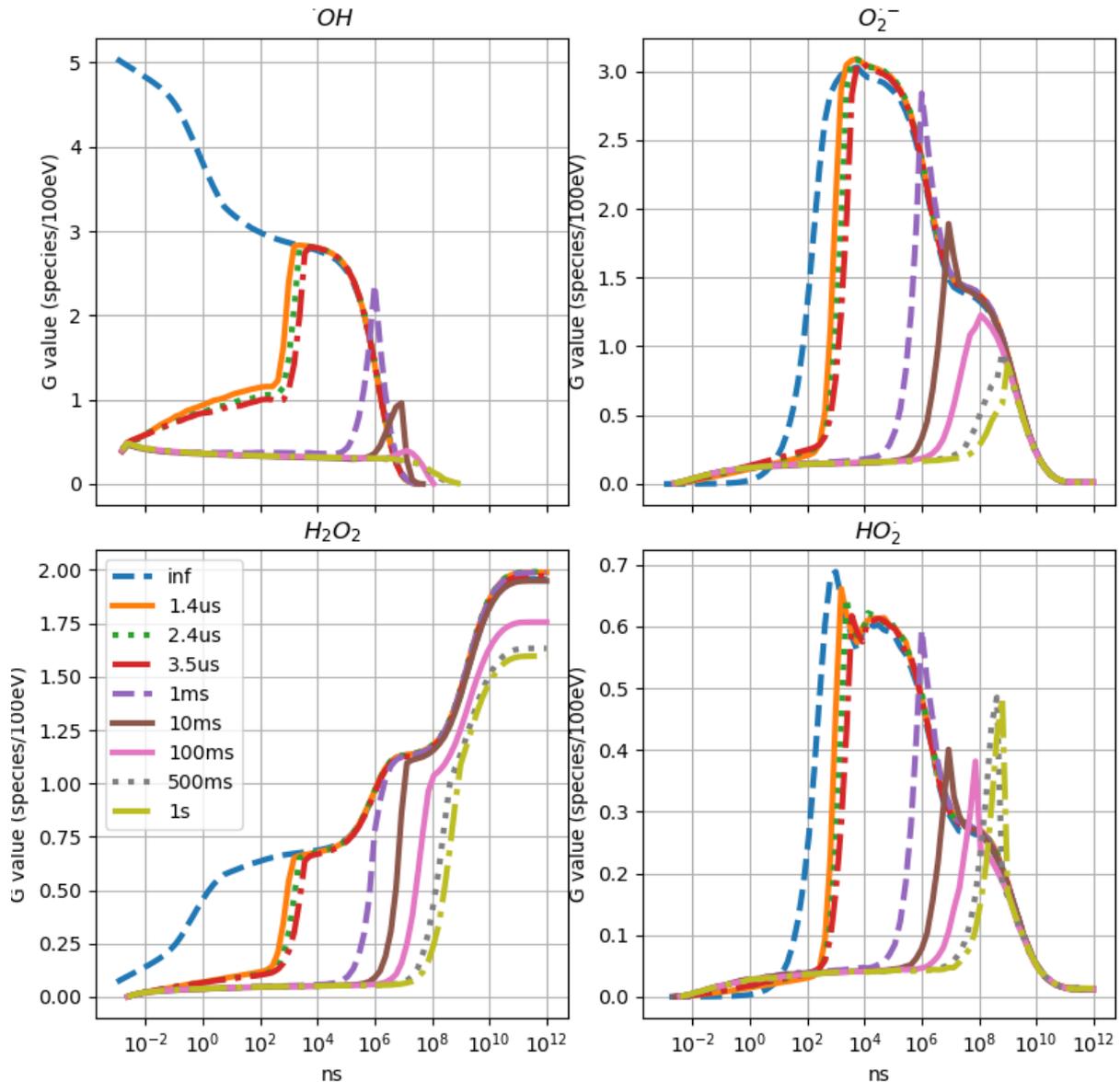

*Figure 9: comparison of the time-dependent G-values of •OH, O₂•⁻, H₂O₂ and HO₂•, obtained from 1 MeV electron beam irradiations into a cubic volume of 3.2 x 3.2 x 3.2 µm³, with 21% oxygenation, up to a total absorbed dose of 0.1 Gy for different pulse durations: infinite (or instantaneous), 1.4 µs, 2.4 µs, 3.5 µs, 1 millisecond (ms), 10 ms, 100 ms, 500 ms, and 1 second (s). Note that G-values are normalized by the total energy deposition during the physical stage.*

The time-dependent G-value is normalized by the total energy deposition during the physical stage. While an instantaneous pulse can simply show the evolution of species through chemical reactions, pulses with a duration (in our case ranging from 1.4 µs to 1 second) reveal the effect of successive integration of species, that happens from the beginning of the chemical stage until when all primary chemical species from the last primary particle are fully integrated into the volume.

Reactive oxygen species ($O_2^{•-}$/$HO_2^{•}$) are rapidly produced under oxygenated conditions through pH equilibrium reactions in the medium. At an oxygen concentration of 21% and a pH of 5.5, a significant number of superoxide radicals ($O_2^{•-}$) are produced and maintained from 100 ns to about 1 s.

As shown in Figure 5, hydrogen peroxide ($H_2O_2$) is mainly produced through two successive processes. The first process involves the recombination of hydroxyl radicals ($^\bullet OH$) due to their high concentrations within the heterogeneous distribution. The second process involves the recombination of superoxide radicals ($O_2^{\bullet -}$) and hydroperoxyl radicals ($HO_2^\bullet$) when these species are generated in sufficiently high concentrations during reactions (1) and (2), primarily through spontaneous disproportionation reactions: $HO_2^\bullet + HO_2^\bullet \rightarrow H_2O_2 + O_2$ and $HO_2^\bullet + O_2^{\bullet -} \rightarrow HO_2^- + O_2$ over a longer evolutionary timescale before being stable at about 15 minutes after irradiation. Note that this duration should be strongly depended on pH of media [33]. At this time scale, $H_2O_2$ is found to be consistent across different time structures with pulse durations ranging from 1.4 µs to 10 ms (Figure 9). However, beyond 10 ms, we interestingly observed a decrease in $H_2O_2$ production up to approximately 18% with a pulse duration of 1 second. By increasing the cumulative dose up to 1 Gy (Appendix: Figure 2) and decreasing it down to 0.01 Gy (Appendix: Figure 3), we observed that the lower $H_2O_2$ levels, which are achieved with longer pulse durations, increase with the higher dose and decrease with the lower dose at 32% and 9%, respectively (Figure 10). This effect will be explained in Section 4 as a "pulse transportation" effect.

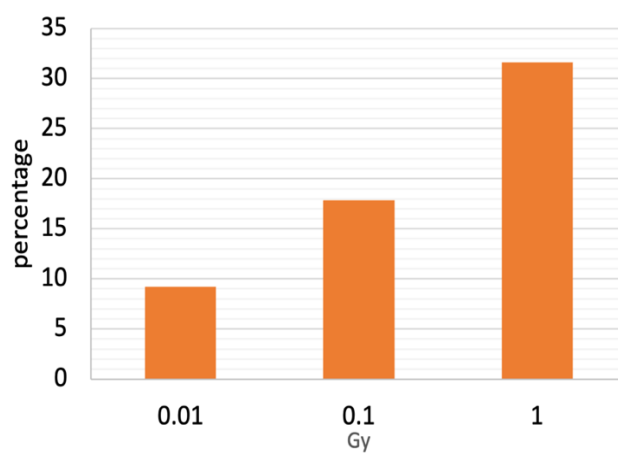

*Figure 10: The percentage of reduced production of $H_2O_2$ (%) as a function of absorbed dose during a pulse duration of 500 milliseconds.*

### 3.3. Impact of multi-pulse time structure

The temporal structure of pulse repetition frequency generated by 1 MeV electrons over a beam duration of 500 ms through a water sample with 21% oxygen concentration is compared with a long single pulse in our simulations. Figure 11-a presents the comparison of the time-dependent G-values of $H_2O_2$ between an infinite pulse of 0.01 Gy, continuous single pulse delivering a dose of approximately 1 Gy (with a dose rate of 2 Gy/s), and multi-pulses with doses of 0.01 Gy (0.02 Gy/s) and 1 Gy (2 Gy/s). For the multi-pulse irradiation, the lower dose rate (0.02 Gy/s) has a DIT of 100 ms (PRF = 10 Hz), and the higher dose rate (2 Gy/s) has a DIT of 10 ms (PRF = 100 Hz).

With a low dose of 0.01 Gy, the multi-pulse structure shows a slightly lower $H_2O_2$ G-value compared to the infinite pulse with the same dose (Figure 11-a). A significant reduction is observed with the higher dose of 1 Gy for both single-pulse and multi-pulse temporal structures, which show very similar $H_2O_2$ production. Note that the high-dose multi-pulse structure has a DIT of 10 ms. For this DIT value, $H_2O_2$ production starts decreasing when the dose rate is increased from 0.02 Gy/s to 4 Gy/s. The decrease in $H_2O_2$ production reaches up to 17.4% at 1 Gy/s and is found to saturate even when the dose rate is increased up to 4 Gy/s.

At the same dose rate of 2 Gy/s, different DITs have varying impacts on $H_2O_2$ production. While a long single pulse results in a lower $H_2O_2$ G-value (1.43 species/100 eV), a longer DIT interval of 100 ms leads to a significantly higher level of $H_2O_2$ production (1.93 species/100 eV) (Figure 11-b). As a result, the reduction in $H_2O_2$ production is no longer observed.

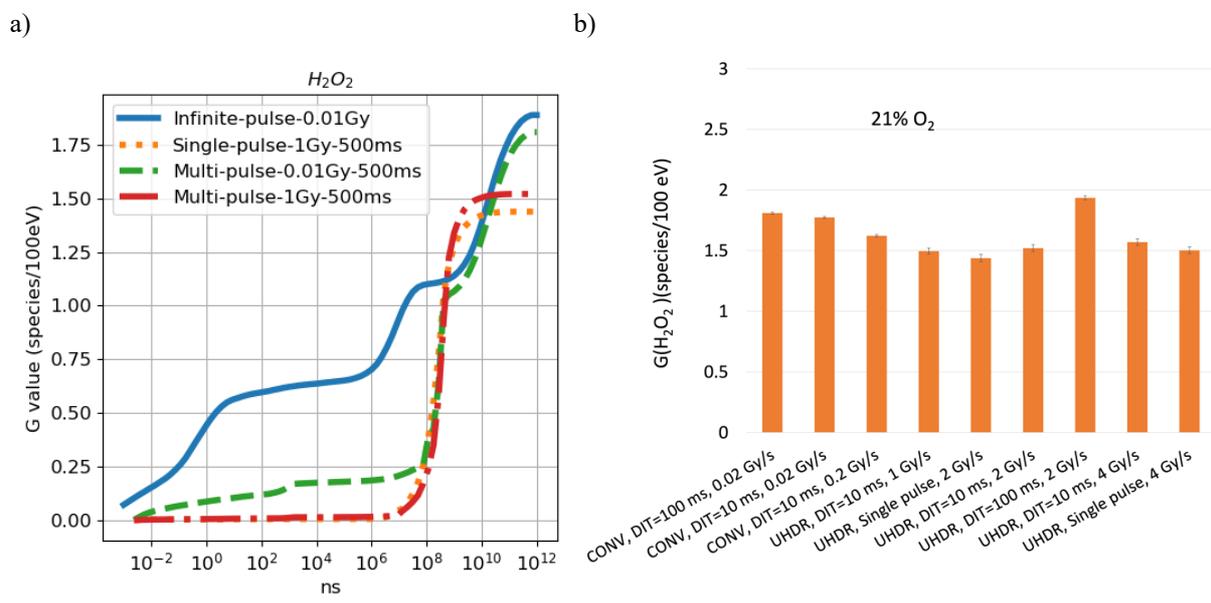

Figure 11: a) Comparison of the time-dependent G-values of $H_2O_2$ obtained by a continuous single pulse of 1 Gy, multi-pulse of doses 0.01 and 1 Gy for over 500 ms beam duration and an infinite pulse for 0.01 Gy through water samples of 21% oxygenation. b) G-values of $H_2O_2$ obtained for different DIT of 10 ms, 100 ms and single pulse at different dose rates of 0.02 Gy/s, 0.2 Gy/s, 1 Gy/s, 2 Gy/s, 4 Gy/s for 500 ms beam duration at 21% oxygenation.

## 4. Discussion

### 4.1. Particle-based model at long time scales.

During the pulse duration, at each delay time, primary chemical species are randomly generated by pre-chemical processes from an incident electron, resulting in a heterogeneous distribution of radiation-induced spurs along the track. This distribution is randomly mixed with the different distribution state of spur evolutions generated from previous incident electrons, resulting in a dynamic period of different track structures integrated along the irradiation time. As a result, this period is very difficult to describe using "well-mixed" models such as the RDME and analytical differential equations. Particle-based models, such as SBS or IRT approaches, are more suitable for describing these states. These models provide spatio-temporal information by simulating the detailed trajectories of individual species, allowing a dynamic account of changes in the spatial distribution and in different evolution times of species in the system. Since pulse durations typically vary from a few microseconds to several seconds, these particle-based models must be able to simulate the evolution of chemical species from the high-concentration species of early radiation-induced spurs to homogeneous kinetics in the bulk solution. The main drawback of the particle-based models is the significant computational time required, especially when simulations involve a large number of species and maintain them for a long time.

Initially introduced by Clifford et al. [15], the IRT model offers a more efficient approach than the SBS one. It assumes that reactions are independent and that the diffusion of reactants from their initial positions to the reaction site is not influenced by other chemical species or volume boundaries. In this context, reaction times for all possible pairs of reactants can be sampled based on reaction probabilities and initial spatial positions, sorted in ascending order of time into a list of potential reactions, and processed independently one by one, starting with the one with the shortest reaction time. This is done without synchronizing the time and position of the involved species. The IRT model is much more computationally efficient than the SBS model, especially when dealing with a large number of species. However, by omitting the computation of diffusion, the spatio-temporal information of the species is not updated during the simulation. Only initial positions and reaction sites are computed.

The synchronous IRT (IRT-syn) [20] model offers an implementation of the IRT approach that provides access to spatio-temporal information at specific intervals, thus offering a solution that is both flexible and efficient. This implementation uses the randomly sampled time, as determined by the IRT model, for the next expected reaction as the time step. In other words, rather than optimizing the time step of several diffusions to coincide with the next reaction, as the SBS model does, the IRT-syn model directly calculates the reaction time using the IRT model. This necessitates synchronizing the time and position of all diffusing species. Based on this new position information, a new list of reactions is recomputed after each time step, which is a drawback of the IRT-syn model. Indeed, to create a new list of reactions, it is necessary to compute the distances of each reaction pair. For a large number of species, if this action is repeated at every time step, a significant amount of time is spent unnecessarily.

The improvements in this work have been made to avoid the calculation of separation distances between pairs of reactants at every time step. At the beginning of the simulation, the list of reactions is computed and added to the event table to be processed (Figure 5). Only dependent events, such as reactions with oxygen, bouncing of chemical species on the volume border, and integrating new primary chemical species through pulse duration, enable a re-computation of the event table. These improvements are essential for handling a large number of species efficiently and for performing simulations over long time scales.

Even though the reaction list is not recomputed at each time step, this implementation maintains synchronization through diffusion at each time step. Thanks to this synchronization algorithm, IRT-syn can provide spatio-temporal information on all species, which can then be coupled with information on geometric boundaries, updated medium changes, or dynamically integrated with new primary chemical species during pulse transportation.

*4.2. Pulse transportation effect*

In water radiolysis under oxygenation conditions, $H_2O_2$ is the final product of ROS pathways. The observation that $H_2O_2$ levels are lower with longer pulse durations compared to shorter pulse durations may offer valuable insights into particle transport during the pulse. Indeed, the chemical interaction between different evolution periods from different tracks of the pulses may be important, in particular under higher dose rates, as they lead to higher concentrations of reactive species compared to lower dose rates. We call this effect, observed in our calculations for pulse durations longer than 10 milliseconds, the "pulse TRANsportation" or "TRAN" effect. *This effect is found when a large ROS-self elimination by chemical interaction between different tracks occurs during an adequate long pulse.* The reactions mainly involved in this effect are shown in Table 6.

| Reaction | Reaction rate |
| --- | --- |

|  | $(L.mol^{-1}.s^{-1})$ |
|---|---|
| $^{\bullet}OH + HO_2^{\bullet} \rightarrow O_2 + H_2O$ | 7.90 x 10$^9$ |
| $^{\bullet}OH + O_2^{\bullet-} \rightarrow O_2 + OH^-$ | 1.07 x 10$^{10}$ |

Table 6: reactions involved in the "TRAN" effect [31].

This effect significantly depends on two pulse features:

- Pulse Duration: The pulse duration should be sufficiently long to allow the primary chemical species in the initial tracks of the pulse to form a high concentration of ROS, specifically $O_2^{\bullet-}/HO_2^{\bullet}$. This equilibrium state of $O_2^{\bullet-}$ and $HO_2^{\bullet}$ is created and maintained for a considerable time (up to 1 second at pH 5.5) through an acid-base reaction equilibrium, which is highly dependent on the pH of the medium [33]. High concentrations of $O_2^{\bullet-}$ and $HO_2^{\bullet}$ promote the reaction of these ROS with $^{\bullet}OH$ radicals in subsequent tracks, thereby competing with the internal hydroxyl radical recombination processes ($^{\bullet}OH + ^{\bullet}OH \rightarrow H_2O_2$) occurring in these tracks.
- Dose Induced in the Pulse: A higher dose rate, resulting in a higher concentration of species, can accelerate the reactions between species. Through the TRAN effect, this higher concentration of species of $O_2^{\bullet-}/HO_2^{\bullet}$ will react with the higher concentration of $^{\bullet}OH$ of later tracks, resulting in a significant drop in ROS. Figure 10 shows the percentage of $H_2O_2$ loss as a function of dose in 500 ms pulse duration. In the same duration, at approximately 1 Gy in the pulse, the TRAN effect reduces $H_2O_2$ by about 32% compared to an 8% reduction at 0.01 Gy.

The higher dose rates mentioned above are defined as the total delivered dose over a sufficiently large period (usually a second), known as the average dose rate. This dose rate differs from the dose per pulse, which is defined as the delivered dose over a short pulse (typically a microsecond) [3]. The simulation results indicate that the TRAN effect, resulting in a lower $H_2O_2$ G-value, depends not on the dose per pulse for short durations ranging from 1.4 µs to 3.5 µs, but on the average dose rate over longer pulses of 10 milliseconds or 100 milliseconds.

However, the TRAN effect does not necessarily occur in a long single pulse but rather in several pulses over the given duration and can be significantly accelerated if these pulses have higher doses or have been delivered in a high repetition frequency, as is the case with UHDR irradiation. Indeed, the next section will discuss the TRAN effect in UHDR and CONV irradiations.

### 4.3. TRAN effect in UHDR and CONV irradiations

Current literature suggests that UHDR delivery exceeds 40 Gy/s, compared to the typical 0.1 Gy/s in CONV radiotherapy. To achieve high dose rates, there may generally be two methods to achieve high dose rates [3]:

- Particle beams are generated by a variable number of pulse trains within a repetition frequency, with UHDR using high repetition frequencies and CONV employing lower repetition frequencies (see Figure 3).
- And (or), UHDR accumulates dose within a higher deliverable dose per pulse via higher beam currents, in contrast to CONV's lower beam currents.

Therefore, over a given time frame, the deliverable dose of UHDR is higher than that of CONV. If this time frame is sufficiently long (about milliseconds), the TRAN effect occurs in both CONV and UHDR, regardless of whether it is in a long single pulse or a temporal structure beam. However, under CONV conditions, the ROS self-elimination effect occurs slowly. In contrast, higher concentrations of reactive species at higher dose rates strengthen the TRAN effect. Consequently, total ROS levels decrease more efficiently through UHDR irradiation than CONV.

In the case of a temporal structure beam, at the same dose rate, a higher repetition frequency of discrete pulse trains enhances the TRAN effect, reaching its maximum when the beam forms a single long pulse. A large interval of DIT (about 100 ms) between consecutive pulses may limit the reactions involved in the TRAN effect (Table 6) between the high concentration of $O_2^{\bullet-}$/$HO_2^\bullet$ from the sooner pulse and $^\bullet OH$ produced by the later pulse. Since $O_2^{\bullet-}$/$HO_2^\bullet$ rapidly decay over 100 ms, the remaining $O_2^{\bullet-}$/$HO_2^\bullet$ are insufficient to sustain the TRAN effect. This DIT dependence is minimal under CONV but significant under UHDR. Note that the decay of $O_2^{\bullet-}$/$HO_2^\bullet$ is strongly dependent on the water's pH level [33] then depends on irradiated media. At higher pH levels, the decay of $O_2^{\bullet-}$/$HO_2^\bullet$ slows down, which may enhance the TRAN effect.

It is interesting to note that C. Ward *et al.* highlighted in [34] lower extracellular pH values in cancer cells than in normal cells. They report extracellular pH values of around 7.4 for normal cells, compared with 6.7-7.1, or even lower, for cancer cells. Based on the simulation results presented in this work, we observed that the TRAN effect is dependent on pH values, demonstrating the importance of taking into account the physico-chemical properties of the irradiated biological medium in simulations. It would therefore be interesting to study experimentally the impact of pH on the TRAN effect.

This reduced production of $H_2O_2$ under UHDR has been experimentally observed in several studies [28,35-40], both under hypoxic and atmospheric oxygen conditions. However, due to the extensive computation time required to simulate these irradiation conditions, a full comparison of time structures between simulation and experiment is beyond the scope of this work.

**5. Conclusion**

A novel hybrid model has been developed that integrates IRT particle-based and RDME approaches to simulate time-structured dose-per-pulse and multi-pulse for water radiolysis at varying absorbed doses. In this work, we presented results for absorbed doses of 0.01, 0.1, 1, and 2 Gy, during beam duration of up to 1 second for a single pulse, and 500 ms for multi-pulse irradiation. The time-structured pulses are derived from realistic raw signals of UHDR pulses and account for beamline characteristics such as temporal pulse duration, pulse repetition frequency, and pulse amplitude. While there is no evidence of a link between the TRAN effect and the FLASH effect, the preliminary simulations indicate that the pulse's time structure can significantly affect the production of ROS in pure water. It is essential to further validate this Geant4-DNA "UHDR" application with experimental data, as it has the potential to help optimize beamline parameters, including dose, timing, and oxygen concentration, to more effectively study ROS levels in water radiolysis.

**Acknowledgment**

L. T. Anh and P. V. Cuong would like to thank the National Foundation for Science and Technology Development of Vietnam (NAFOSTED) for its financial support under Grant No. 103.04-2023.103. The authors also gratefully acknowledge Amentum Pty Ltd for granting permission to redistribute the original open-source version of PBC in Geant4. Additionally, the authors would like to thank Dr. Mathieu Karamitros and Dr. Ianik Plante for their valuable discussions. The authors are also grateful to L. X. Chung (INST,VINATOM), D. T. K. Linh (INST,VINATOM), and B. T. Hung (Phenika) for their generous supports for this work.

**Appendix:**

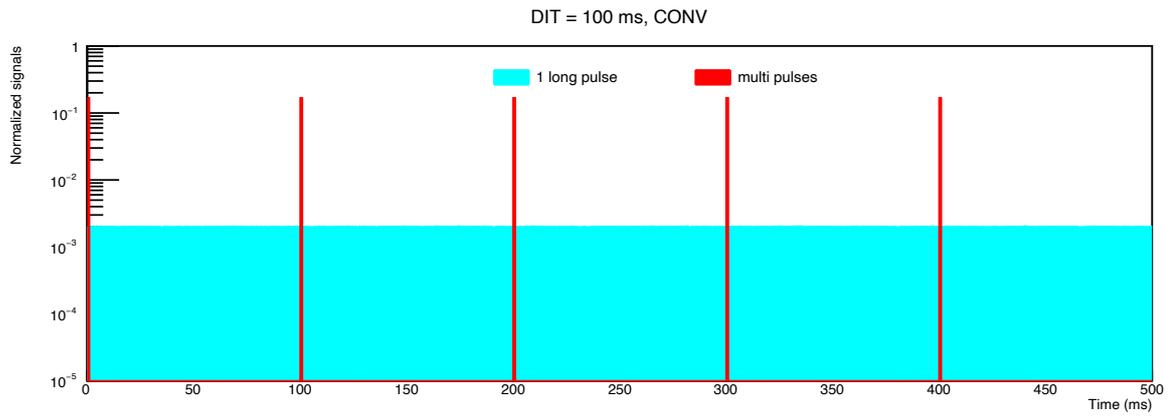

*Figure 1: An example of beam time structure used in the simulation of 5 pulses (red lines) with DIT of 100 ms and a long single pulse (blue rectangle) over 500 milliseconds.*

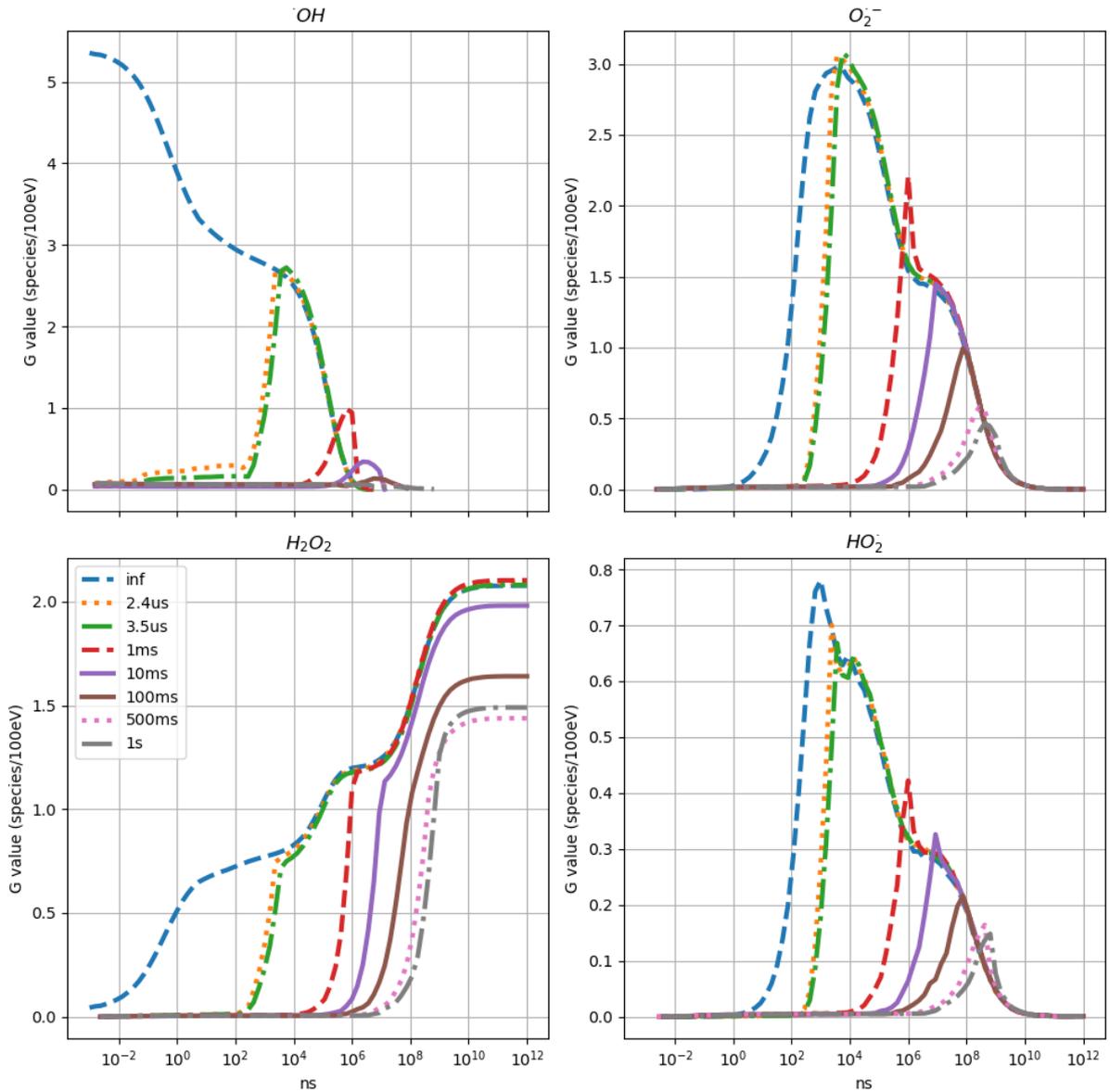

*Figure 2: Comparison of the time-dependent G-values of •OH, $O_2^{•-}$, $H_2O_2$ and $HO_2^{•}$, obtained from 1 MeV electron beam irradiations into a cubic volume of 1.6 x 1.6 x 1.6 µm³, with **21% oxygenation**, up to a total absorbed dose of 1 Gy for different pulse durations: infinite, 2.4 µs, 3.5 µs, 1 millisecond (ms), 10 ms, 100 ms, 500 ms, and 1 second (s), corresponding to dose rates ranging from 1 Gy/s to $7.14 \times 10^5$ Gy/s.*

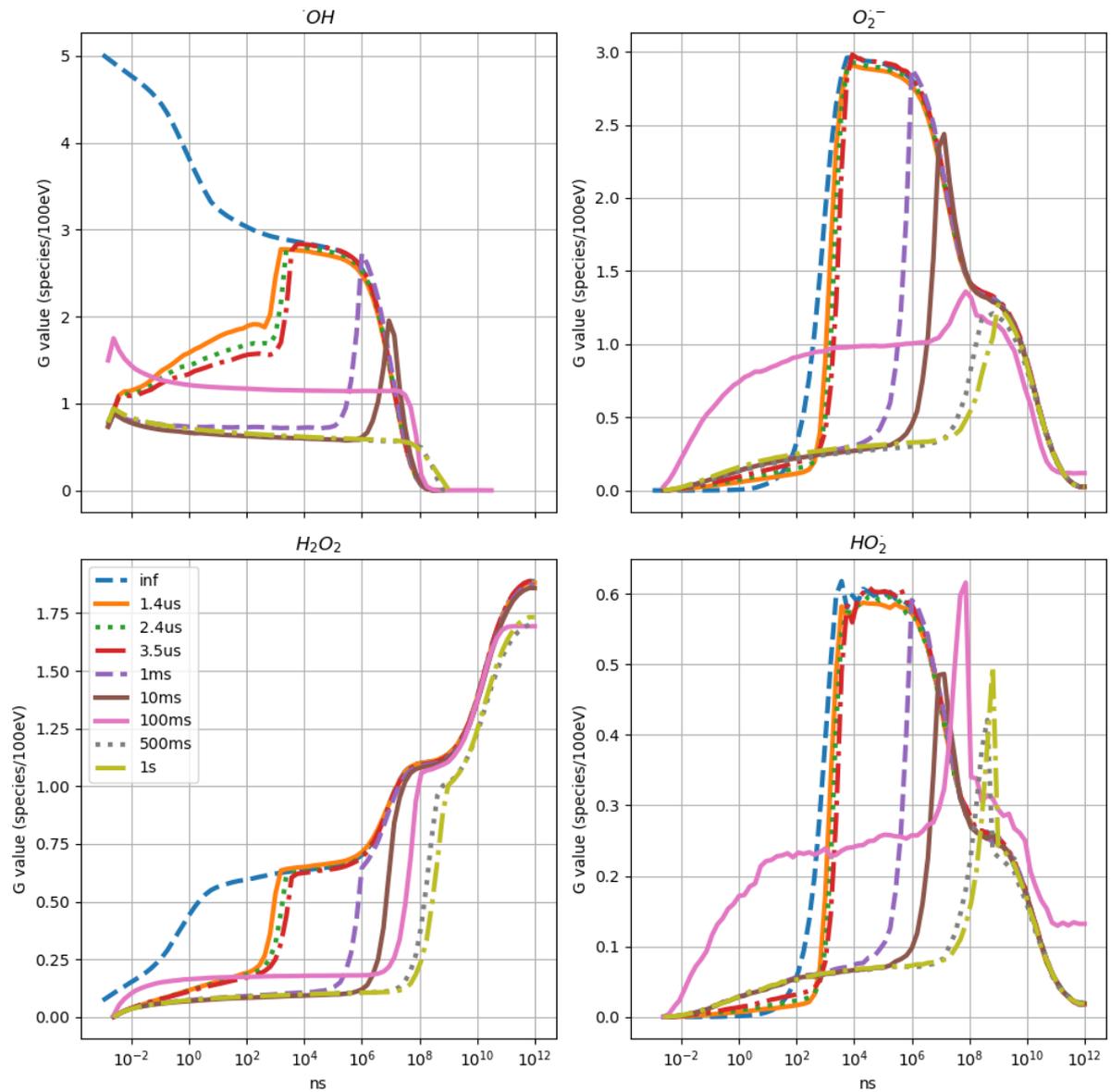

*Figure 3: Comparison of the time-dependent G-values of •OH, $O_2$•⁻, $H_2O_2$ and $HO_2$• obtained from 1 MeV electron beam irradiations into a cubic volume of 3.2 x 3.2 x 3.2 µm³, with **21% oxygenation**, up to a total absorbed dose of 0.01 Gy for different pulse durations: infinite, 1.4 µs, 2.4 µs, 3.5 µs, 1 millisecond (ms), 10 ms, 100 ms, 500 ms, and 1 second (s), corresponding to dose rates ranging from 0.01 Gy/s to 7.14 × 10³ Gy/s.*

1